# METHOD OF LINEAR INVARIANTS FOR DESCRIPTION OF BEAM DYNAMICS OF FEL UNDULATOR

A. Angelow [1]*, D. Trifonov[2], V. Angelov[2], H. Hristov[3]

[1]72 Trackia Blvd., 1784 Sofia, Institute of Solid State Physics, BAS
[2]Institute for Nuclear Research and Nuclear Energy, BAS
[3]Plovdiv University, Faculty of Physics, Department of Nuclear Physics

**Abstract** We propose a new model for description of electrons beam dynamics in Free Electron Laser (FEL) undulator, based on the method of linear time-dependent invariants of quantum-mechanical charge particle. The magnetic field has periodic structure along the undulator. For this problem, described by time-dependent quadratic Hamiltonian, we obtain exact solution. The time-evolutions of the tree quantum fluctuations: covariance cov(q,p), var(q) and var(p) for the charge particle in this case are also determined. This research will help to optimize the FEL undulator: for example, using a 2.5 GeV linear electron accelerator it will be possible to emit radiation at 1.5 nm and shorter length. The radiation of this range is a power tool for investigation and innovative technologies development in the field of materials science, biology, medicine, non linear optics, x-ray microscopy; proteine crystallography and even to grow carbon nanotubes with high productivity.
This method could be applicable also to any device with periodic structure of applied field (e.g. Tokamak, cyclic accelerators) for the case of charge non-relativistic quantum particles.
**Keywords:** Quantum Harmonic Oscillator, Schrödinger Uncertainty Relation, Linear Integrals of Motion, Free Electron Laser, undulator.

## Introduction

The problem of creation of shortwave coherent radiation sources in general aspects reduces to implementation of free electron lasers. Various schemes of X-ray FELs have been considered [1]. Understanding the effects of the undulator is of critical importance in the design of an x-ray FEL [2], and up to now quantum mechanical consideration of the electron motion in the undulator has not been done.

It has been shown [3], that a motion of quantum particle in a time-dependent magnetic or electric field can be treated in a similar way as in the case of a Nonstationary Quantum Harmonic Oscillator (NQHO). Our approach in solving the problem of charged particles moving in an undulator is based completely on QM, namely on the base of linear integrals of motion for quantum particle with quadratic Hamiltonians to find the exact solutions of such equations [3], and to express all the thee second order central moments [4,5].

## Method of linear integrals of motion of electron in undulator

For the first time the motion of charge quantum particle in **constant** homogeneous magnetic field was considered by Landau in 1930 (see, for ex. [6]). In [3] the motion of charge particles in arbitrary **time-dependent** homogeneous electromagnetic field has been investigated, and we intend to apply a modification of this method to the beam dynamics of an **undulator**, Fig. 1.

This study can be viewed as a first attempt to apply quantum invariants method to the case of non-homogenious field. Here we split the space in regions for each of them we easily find quantum solutions. We consider an idealization of the undulator, consisted of periodically placed magnets, with alternative changing the direction of constant magnetic field in z direction, and zero between magnets, Fig. 1. It is worth noting that the regions are pick out so narrow, that the electrons deflect slightly in y-direction, and keep emitted X-rays collimated.

---
[1] E-mail:<a_angelow@phys.bas.bg>

**First region** (charged particle in a constant field *H*): The linear invariants in first region are constructed on the basis of [3]:

$$B^I = \sqrt{\frac{m\omega}{2}} e^{i\omega t}(x_0 - iy_0), \quad A^I = \sqrt{\frac{m\omega}{2}} e^{i\omega t}[y - y_0 - i(x - x_0)], \quad (1)$$

where *m* is the particle mass, $\omega$ is the cyclotron frequency, and $x_0, y_0$ are the well known operators for 'coordinates' of the center of the orbit of a charged particle in a constant magnetic field *H* (we use the system $\hbar = c = 1$):

$$x_0 = \frac{1}{2}x + \frac{1}{m\omega}p_y, \quad y_0 = \frac{1}{2}y - \frac{1}{m\omega}p_x, \quad \omega = eH/m. \quad (2)$$

Actually, the 'trajectory' is very small part of circle orbit, and the mean values of $x_0, y_0$ are different in the third region where the magnetic field is in the opposite direction: $<y_0>$ is alternating along an axis parallel to X, while $<x_0>$ accumulates its value periodically. The $\psi$-function in the first region is [3]:

$$\psi^I_{\alpha\beta}(x,y,t) = \sqrt{\frac{m\omega}{2\pi}} \exp[-\frac{i}{2}m\omega t - \frac{1}{4}m\omega(x^2+y^2)].$$

$$\cdot \exp[-\frac{1}{2}(|\alpha|^2+|\beta|^2) + \sqrt{\frac{m\omega}{2\pi}}(\beta(x+iy)+i\alpha(x-iy)e^{-i\omega t})-i\alpha\beta e^{-i\omega t}]. \quad (3)$$

This wave function is an eigenstate of both invariant operators $A^I$ and $B^I$:

$$A^I \psi^I_{\alpha\beta} = \alpha \psi^I_{\alpha\beta}, \qquad B^I \psi^I_{\alpha\beta} = \beta \psi^I_{\alpha\beta}. \quad (4)$$

The above invariants and wave function are particular cases of the two non-Hermitian linear invariants and their eigenstates (coherent states (CS)), constructed in [3] for a charged particle motion in a time dependent magnetic field *H(t)*. The invariants and CS in [3] are expressed in terms of a function $\varepsilon(t)$, that obeys the classical oscillator equation of motion $\ddot{\varepsilon}(t) + \Omega^2(t)\varepsilon(t) = 0$, $\Omega(t) = eH(t)/m$, with Wronskian $\dot{\varepsilon}\varepsilon^* - \varepsilon^*\dot{\varepsilon} = 2ie/m$. The above provided solutions for a constant magnetic field *H* correspond to the choice $\varepsilon(t) = \sqrt{2c/H} \exp(i\omega t/2) \equiv \varepsilon^I(t)$.

**Second region** (free particle, *H=0*). The linear invariants in the second region are constructed on the basis of [3]:

$$A^{II} = \frac{1}{2\sqrt{e}}[(iC_1 t + \frac{e}{mC_1})(p_x + ip_y) + mC_1(y - ix)],$$

$$B^{II} = \frac{1}{2\sqrt{e}}[(iC_1 t + \frac{e}{mC_1})(p_y + ip_x) + mC_1(x - iy)], \quad (5)$$

where $C_1$ is an arbitrary positive constant. The $\psi$-function in the second region is [3]:

$$\psi''_{\alpha\beta}(x,y,t) = \sqrt{\frac{e}{\pi}} \frac{1}{iC_1 t + \frac{e}{mC_1}} \exp[-\frac{mC_1}{2(iC_1 + \frac{e}{mC_1})}(x^2+y^2)]. \quad (6)$$

$$.\exp[-\frac{1}{2}(|\alpha|^2+|\beta|^2) + \frac{\sqrt{e}}{\sqrt{C_1^2 t^2 + \frac{e^2}{m^2 C_1^2}}}(\beta(x+iy)e^{-i\gamma(t)} + \alpha(x-iy)e^{-i\gamma(t)}) - i\alpha\beta e^{-i2\gamma(t)}]$$

where $\gamma(t) = (e/m)\int_0^t \frac{d\tau}{C_1^2\tau^2 + \frac{e^2}{m^2 C_1^2}} = \arctan(mC_1^2 t/e) + C_2$.

The above invariants and wave function in the second region also are particular cases of the two non-Hermitian linear invariants and their eigenstates, constructed in [3], this time corresponding to the choice $\varepsilon(t) = iC_1 t + e/mC_1 \equiv \varepsilon''(t)$.

## The mean trajectory. Evolution of the first and second statistical moments

**First region** (constant field $H$). We express the coordinate and momentum operators in terms of the invariants $A^I$, $B^I$ and their conjugate [3]

$$x = \sqrt{e}[B^I + ie^{-i\omega t}A^I + (B^I)^+ - ie^{i\omega t}(A^I)^+], \quad y = \sqrt{e}[iB^I + e^{-i\omega t}A^I - i(B^I)^+ + e^{i\omega t}(A^I)^+]$$

$$p_x = -\frac{m\omega\sqrt{e}}{2}[iB^I - e^{-i\omega t}A^I - i(B^I)^+ - e^{i\omega t}(A^I)^+], \quad (7)$$

$$p_y = -\frac{m\omega\sqrt{e}}{2}[-B^I + ie^{-i\omega t}A^I - (B^I)^+ - ie^{i\omega t}(A^I)^+].$$

Using the eigenvalue equations for $A^I$, $B^I$ we easily obtain the mean coordinates and mean moments in states $\psi^I_{\alpha\beta}$,

$$<x>_I = \sqrt{e}[\beta^I + ie^{-i\omega t}\alpha^I + \beta^{I*} - ie^{i\omega t}\alpha^{I*}] \equiv \bar{x}_I(t),$$
$$<y>_I = \sqrt{e}[i\beta^I + e^{-i\omega t}\alpha^I - i\beta^{I*} + e^{i\omega t}\alpha^{I*}] \equiv \bar{y}_I(t), \quad (8)$$

It is clear from these mean coordinates that the mean trajectory is a circle in the *X-Y* plane with radius $r = 2|\alpha^I|\sqrt{e}$ centered at point with coordinates

$$X_0 = 2\sqrt{e}\,\text{Re}\,\beta^I, \qquad Y_0 = -2\sqrt{e}\,\text{Im}\,\beta^I \quad (9)$$

$\bar{x}_I(t) = 2\sqrt{e}\,\text{Re}\,\beta^I + 2|\alpha^I|\sqrt{e}\sin(\omega t - \varphi_\alpha), \quad \bar{y}_I(t) = -2\sqrt{e}\,\text{Im}\,\beta^I + 2|\alpha^I|\sqrt{e}\cos(\omega t - \varphi_\alpha)$

*The second statistical moments* (variances [4] and covariances [5]) of coordinates and kinetic moments can be easily calculated [3], using the above expressions of $x, y$ and $p_x, p_y$ in terms of the invariants $A, B$ and their conjugate. The result is (here we restored $\hbar$ for clarity):

$$(\Delta x^I)^2 := <xx> - <x>^2 = \frac{\hbar}{m\omega} = (\Delta y^I)^2, \qquad (\Delta p_x^I)^2 := <p_x p_x> - <p_x>^2 = \frac{m\omega\hbar}{4} = (\Delta p_y^I)^2$$

$$\text{cov}(x^I, p_x^I) := \frac{1}{2}<xp_x + p_x x> - <x><p_x> = 0 = \text{cov}(y^I, p_y^I). \tag{10}$$

We see that the second moments are all constant in time and minimize the Heisenberg uncertainty relation for coordinates and moments. The covariances are vanishing, thus the second moments minimize the more precise Schrödinger uncertainty inequality.

**Second region** (free particle, $H=0$). The expressions of coordinate and momentum operators in terms of the non-Hermitian invariants for the free particle take the form

$$x = \frac{\sqrt{e}}{2mC_1}[B^{II} + (B^{II})^+ + i(A^{II} - (A^{II})^+) - i\frac{mC_1^2}{e}(B^{II} - (B^{II})^+) + \frac{mC_1^2}{e}(A^{II} + (A^{II})^+)],$$

$$y = \frac{\sqrt{e}}{2mC_1}[i(B^{II} - (B^{II})^+) + A^{II} + (A^{II})^+ - \frac{e}{mC_1 t}(B^{II} + (B^{II})^+) + i\frac{e}{mC_1 t}(A^{II} - (A^{II})^+)],$$

(11)

$$p_x = \frac{mC_1}{2\sqrt{e}}[-i(B^{II} - (B^{II})^+) + A^{II} + (A^{II})^+], \quad p_y = \frac{\sqrt{e}}{2C_1 t}[B^{II} + (B^{II})^+) - i(A^{II} - (A^{II})^+)]$$

Then, using the eigenalue equations for the invariants $A^{II}, B^{II}$ we get the mean coordinates in $\psi^{II}{}_{\alpha\beta}$

$$<x>_{II} = \frac{\sqrt{e}}{2mC_1}[\beta^{II} + \beta^{II*} + i(\alpha^{II} - \alpha^{II*}) - \frac{mC_1^2}{e}(i(\beta^{II} - B^{II*}) - (\alpha^{II} + \alpha^{II*}))t] \equiv \bar{x}_{II}(t),$$

$$<y>_{II} = \frac{\sqrt{e}}{2mC_1}[i(\beta^{II} - \beta^{II*}) + \alpha^{II} + \alpha^{II*} - \frac{e}{mC_1}((\beta^{II} + \beta^{II*}) - i(\alpha^{II} - \alpha^{II*}))\frac{1}{t}] \equiv \bar{y}_{II}(t).$$

$$\bar{p}_x^{II} = \frac{mC_1}{\sqrt{e}}[\text{Im}\,\beta^{II} + \text{Re}\,\alpha^{II}], \qquad \bar{p}_y^{II} = \frac{\sqrt{e}}{C_1 t}[\text{Re}\,\beta^{II} + \text{Im}\,\alpha^{II}] \tag{12}$$

For the second moments in the second region (free particle in CS $\psi^{II}{}_{\alpha\beta}$) we find the expressions

$$(\Delta x^{II})^2 := <xx> - <x>^2 = \frac{|e|}{2m^2 C_1^2} + \frac{C_1^2}{2|e|}(t-t_1)^2 = (\Delta y^{II})^2,$$

$$(\Delta p_x^{II})^2 := <p_x p_x> - <p_x>^2 = \frac{m^2 C_1^2}{2|e|} = (\Delta p_y^{II})^2$$

$$\text{cov}(x, p_x) := \frac{1}{2}<xp_x + p_x x> - <x><p_x> = -\frac{mC_1^2}{2e}(t-t_1) = \text{cov}(y, p_y) \tag{13}$$

where $t_1$ is the initial time for the motion in the second region. We see that the coordinate variances and the coordinate-moment covariances increase in time (the wave packet is spreading), while the variances of the two moments are constant. The Heisenberg uncertainty relation is not minimized in these Covariance States (CovS), while the more précised Schrödinger uncertainty relation [7]

$$(\Delta x)^2 (\Delta p_x)^2 - \text{cov}^2(x, p_x) \geq \hbar^2/4 \tag{14}$$

is minimized, due to the nonvanishing covariance $\text{cov}(x, p_x)$, similar to CovS in [8].

**Stitching together the two trajectories.** Let at the initial moment $t = 0$ the particle enters the first region with velocity parallel to X-axis, and let the first region is a strip in the X-Y plane with width $(x = 0, x = x_1)$, and the second region be with X-coordinates $(x_1, x_2)$. This means that we choose $\beta^I = 0, \alpha^I > 0$, and $\bar{x}_I(0) = 0$, $\bar{y}_I(0) = 2\alpha^I \sqrt{e} \equiv y_0$. Denote by $t_1$ the moment when $\bar{x}_I(t_1) = x_1$. Since we want the particle motion to alternate along the a line parallel to X-axis we put $\bar{y}_I(t_1) = y_0/2 = \alpha^I \sqrt{e}$, i.e. $t_1 = \arccos(1/2)/\omega$.
The final conditions of the first region have to be set as initial conditions for the trajectory in the second region, i.e.
$$\bar{x}_I(t_1) = \bar{x}_{II}(t_1), \quad \bar{y}_I(t_1) = \bar{y}_{II}(t_1); \quad \bar{p}_x^I(t_1) = \bar{p}_x^{II}(t_1), \quad \bar{p}_y^I(t_1) = \bar{p}_y^{II}(t_1).$$
From these equations we get that the following four relations have to be valid

$$2\alpha^I \sin(\arccos(1/2)) = \frac{1}{mC_1}[\text{Re}\,\beta^{II} - \text{Im}\,\alpha^{II} + \frac{mC_1^2}{e\omega}(\text{Re}\,\alpha^{II} + \text{Im}\,\beta^{II})\arccos(1/2)],$$
(15)
$$\alpha^I = \frac{1}{mC_1}[\text{Re}\,\alpha^{II} - \text{Im}\,\beta^{II} - \frac{e\omega}{mC_1}(\text{Re}\,\beta^{II} + \text{Im}\,\alpha^{II})/\arccos(1/2)].$$

$$\frac{\omega\alpha^I}{\sqrt{2}} = \frac{C_1}{\sqrt{e}}[\text{Re}\,\alpha^{II} - \text{Im}\,\beta^{II}], \quad \alpha^I \sin(\arccos(1/2)) = \frac{-1}{C_1 \arccos(1/2)}[\text{Re}\,\beta^{II} + \text{Im}\,\alpha^{II}]$$

We see that the two mean trajectories are stitched together for any $\alpha^I$ if $\alpha^{II}$, $\beta^{II}$ and $C_1$ are related according to the following two equations (noted that $\arccos(1/2) = \pi/3$)

$$\frac{\sin(\pi/3)}{2mC_1}[\text{Re}\,\beta^{II} - \text{Im}\,\alpha^{II} + \frac{\pi mC_1^2}{3e\omega}(\text{Re}\,\alpha^{II} + \text{Im}\,\beta^{II})]$$
$$= \frac{1}{mC_1}[\text{Re}\,\alpha^{II} - \text{Im}\,\beta^{II} - \frac{3e\omega}{\pi mC_1}(\text{Re}\,\beta^{II} + \text{Im}\,\alpha^{II})],$$

$$\frac{C_1\sqrt{2}}{\omega\sqrt{e}}[\text{Re}\,\alpha^{II} - \text{Im}\,\beta^{II}] = -\frac{3(\text{Re}\,\beta^{II} + \text{Im}\,\alpha^{II})}{\pi C_1 \sin(\pi/3)}. \tag{16}$$

Evidently these conditions can be always satisfied since $\alpha^{II}$ and $\beta^{II}$ are still free. Next we have to stitch together all the second moments of coordinates and moments, which are provided in the above. One can check that for this purpose is sufficient it to fix the parameter $C_1$ as follows: $C_1 = \sqrt{|e|\omega/2m}$. The stitching procedure described above, can be applied to other next neighbor regions of the undulator. In this way we obtain the quantum mean trajectory oscillating along the axis $y = y_0/2$, parallel to X-axis.

## Conclusion

We have shown that charged particle motion in an idealized undulator (in which the alternation of the magnetic field along the undulator axis is modeled by a step function) can be described in the framework of quantum mechanics. The classical trajectory of the particles coincides with the mean quantum trajectory in the coherent states, which are chosen as eigenstates of two linear non-Hermitian invariants. These wave packets possess the advantage of minimal quantum mechanical fluctuations in sense of Schrödinger uncertainty inequality – in both type of regions in the above chosen states the calculated three second moments of coordinates and moments do minimize the Schrödinger relation.

This study can be regarded as a first attempt of quantum invariants treatment of beam dynamics for the more realistic case of particle motion in a non-homogenious field.

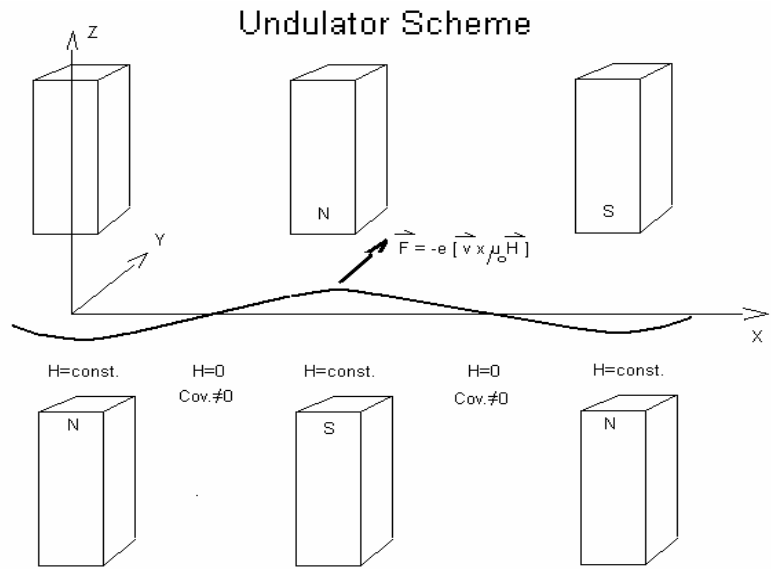

Figure 1. The motion of electrons is in X-Y plane. The X-rays radiation is along X- axis